\documentclass[a4paper,twocolumn]{article}
\usepackage{txfonts}

\usepackage{moreverb,url}

\usepackage[colorlinks,bookmarksopen,bookmarksnumbered,citecolor=red,urlcolor=red]{hyperref}

\newcommand\BibTeX{{\rmfamily B\kern-.05em \textsc{i\kern-.025em b}\kern-.08em
T\kern-.1667em\lower.7ex\hbox{E}\kern-.125emX}}

\usepackage{graphicx}
\usepackage{program}
\usepackage{ifthen}
\usepackage{epsfig}
\usepackage{array}
\usepackage{color}
\usepackage[table]{xcolor}
\usepackage{xspace}
\usepackage{listings}
\usepackage{wasysym}
\usepackage{booktabs}
\usepackage{newfloat}
\usepackage{multirow}
\usepackage[labelfont=bf,labelsep=period,font={small}]{caption}
\DeclareFloatingEnvironment{algorithm}

\definecolor{darkgreen}{rgb}{0,0.44,0}
\definecolor{darkred}{rgb}{0.44,0,0}
\definecolor{darkblue}{rgb}{0,0,0.44}
\definecolor{mygray}{rgb}{0.9,0.9,0.9}
\definecolor{mymauve}{rgb}{0.58,0,0.82}
\definecolor{myred}{rgb}{0.72,0.18,0.0} 
\definecolor{mygreen}{rgb}    {0.0,0.72,0.0} 
\definecolor{myblue}{rgb} {0.18,0.0,0.72} 
\definecolor{mycreme}{rgb}        {1.0,0.8,0.2} 

\usepackage{arydshln}

\newcolumntype{I}{!{\vrule width 1.5pt}}

\newlength\savedwidth 

\newcommand\whline{\noalign{\global\savedwidth\arrayrulewidth 
                            \global\arrayrulewidth 1.5pt}%
           \hline 
           \noalign{\global\arrayrulewidth\savedwidth}}

\newcommand{\routinename}{}

\newcommand{\precondition}{~}

\newcommand{\postcondition}{~}

\newcommand{\invariant}{~}

\newcommand{\guard}{~}

\newcommand{\partitionings}{~}

\newcommand{\partitionsizes}{~}

\newcommand{\blocksize}{blank}

\newcommand{\repartitionings}{~}

\newcommand{\repartitionsizes}{~}

\newcommand{\moveboundaries}{~}

\newcommand{\beforeupdate}{~}

\newcommand{\afterupdate}{~}

\newcommand{\NoShow}[1]{}

\newcommand{\FlaAlgorithm}{
\begin{tabular}{| p{0.92\textwidth}|} \hline
$\mbox{\color{blue}Algorithm:~}\routinename$
\\ \whline
\partitionings \\
$\mbox{\color{blue} ~~~where~}$ \partitionsizes 
\\ 
$\mbox{\color{blue}while~} \guard \mbox{~\color{blue} do}$
\\
\ifthenelse{\equal{\blocksize}{1}}{\\}%
{%
\ifthenelse{ \equal{\blocksize}{blank} }{}%
{~~~~{\bf Determine block size $ \blocksize $}\\}%
}
~~~~ 
\repartitionings \\
~~~$\mbox{\color{blue} ~~~where~}$ \repartitionsizes
\\ \hline
~~~~  \update 
\\ \hline
~~~~ 
\moveboundaries 
\\
$\mbox{\color{blue} endwhile} $
\\ \hline 
\end{tabular}
}

\newcommand{\FlaWorksheet}{
\begin{tabular}{| c | p{0.98\textwidth} |}\hline
Step & $\mbox{\color{blue}Algorithm:~}\routinename$
\\ \hline
\rowcolor{yellow!75}
1a & $ \precondition $ 
\\ \whline
4 & 
\begin{minipage}[t]{0.9\textwidth}%
\partitionings~ \\
$\mbox{\color{blue} ~~~where~}$ \partitionsizes
\end{minipage}
\\ \hline
\rowcolor{yellow!75}
2 & $ \invariant $ 
\\ \hline
3 &$\mbox{\color{blue}while~} \guard \mbox{~\color{blue} do}$
\\ \hline 
\rowcolor{yellow!75}
2,3 & ~~~~ $ \invariant \wedge \guard$ 
\\ \hline
5a & ~~~~ \begin{minipage}[t]{0.85\textwidth}%
\ifthenelse{\equal{\blocksize}{1}}{}%
{%
\ifthenelse{ \equal{\blocksize}{blank} }{}%
{{\bf Determine block size $ \blocksize $}\\}%
}
\repartitionings~ \\
$\mbox{\color{blue} ~~~where~}$ \repartitionsizes
\end{minipage}
\\ \hline
\rowcolor{yellow!75}
6 & ~~~~ $\beforeupdate $
\\ \hline
8 & ~~~~  \update 
\\ \hline
5b & ~~~~ \begin{minipage}[t]{0.85\textwidth}%
\moveboundaries~
\end{minipage}
\\ \hline
\rowcolor{yellow!75}
7 & ~~~~ $\afterupdate $
\\ \hline
\rowcolor{yellow!75}
2 & ~~~~ $ \invariant  $ 
\\ \hline
 &$\mbox{\color{blue} endwhile} $
\\ \hline \whline
\rowcolor{yellow!75}
2,3 & $ \invariant \wedge \neg( \guard )$ 
\\ \hline
\rowcolor{yellow!75}
1b & $ \postcondition $ 
\\ \hline
\end{tabular}
}

\newcommand{\FlaWorksheetNine}{
\begin{tabular}{| c | p{0.98\textwidth} |}\hline
Step & $\mbox{\color{blue}Algorithm:~}\routinename$
\\ \hline
\rowcolor{yellow!75}
\phantom{1a} & $ \phantom\precondition $ 
\\ \whline
\phantom{4} & 
\begin{minipage}[t]{0.9\textwidth}%
\partitionings~ \\
$\mbox{\color{blue} ~~~where~}$ \partitionsizes
\end{minipage}
\\ \hline
\rowcolor{yellow!75}
\phantom{2} & $ \phantom\invariant $ 
\\ \hline
\phantom{3} &$\mbox{\color{blue}while~} \guard \mbox{~\color{blue} do}$
\\ \hline 
\rowcolor{yellow!75}
\phantom{2,3} & ~~~~ $ \phantom\invariant \phantom \wedge \phantom
                \guard $ 
\\ \hline
\phantom{5a} & ~~~~ \begin{minipage}[t]{0.85\textwidth}%
\ifthenelse{\equal{\blocksize}{1}}{}%
{%
\ifthenelse{ \equal{\blocksize}{blank} }{}%
{{\bf Determine block size $ \blocksize $}\\}%
}
\repartitionings~ \\
$\mbox{\color{blue} ~~~where~}$ \repartitionsizes
\end{minipage}
\\ \hline
\rowcolor{yellow!75}
\phantom{6} & ~~~~ $\phantom\beforeupdate $
\\ \hline
\phantom{8} & ~~~~  \update 
\\ \hline
\phantom{5b} & ~~~~ \begin{minipage}[t]{0.85\textwidth}%
\moveboundaries~
\end{minipage}
\\ \hline
\rowcolor{yellow!75}
\phantom{7} & ~~~~ $\phantom\afterupdate $
\\ \hline
\rowcolor{yellow!75}
\phantom{2} & ~~~~ $ \phantom\invariant  $ 
\\ \hline
 &$\mbox{\color{blue} endwhile} $
\\ \hline \whline
\rowcolor{yellow!75}
\phantom{2,3} & $ \phantom\invariant \wedge \neg( \phantom\guard )$ 
\\ \hline
\rowcolor{yellow!75}
\phantom{1b} & $ \phantom\postcondition $ 
\\ \hline
\end{tabular}
}

\newcommand{\FlaWorksheetEight}{
\begin{tabular}{| c | p{0.98\textwidth} |}\hline
Step & $\mbox{\color{blue}Algorithm:~}\routinename$
\\ \hline
\rowcolor{yellow!75}
1a & $ \precondition $ 
\\ \whline
4 & 
\begin{minipage}[t]{0.9\textwidth}%
\partitionings~ \\
$\mbox{\color{blue} ~~~where~}$ \partitionsizes
\end{minipage}
\\ \hline
\rowcolor{yellow!75}
2 & $ \invariant $ 
\\ \hline
3 &$\mbox{\color{blue}while~} \guard \mbox{~\color{blue} do}$
\\ \hline 
\rowcolor{yellow!75}
2,3 & ~~~~ $ \invariant \wedge \guard $ 
\\ \hline
5a & ~~~~ \begin{minipage}[t]{0.85\textwidth}%
\ifthenelse{\equal{\blocksize}{1}}{}%
{%
\ifthenelse{ \equal{\blocksize}{blank} }{}%
{{\bf Determine block size $ \blocksize $}\\}%
}
\repartitionings~ \\
$\mbox{\color{blue} ~~~where~}$ \repartitionsizes
\end{minipage}
\\ \hline
\rowcolor{yellow!75}
6 & ~~~~ $\beforeupdate $
\\ \hline
\rowcolor{orange!50}    
8 & ~~~~  \update 
\\ \hline
5b & ~~~~ \begin{minipage}[t]{0.85\textwidth}%
\moveboundaries~
\end{minipage}
\\ \hline
\rowcolor{yellow!75}
7 & ~~~~ $\afterupdate $
\\ \hline
\rowcolor{yellow!75}
2 & ~~~~ $ \invariant  $ 
\\ \hline
 &$\mbox{\color{blue} endwhile} $
\\ \hline \whline
\rowcolor{yellow!75}
2,3 & $ \invariant \wedge \neg( \guard )$ 
\\ \hline
\rowcolor{yellow!75}
1b & $ \postcondition $ 
\\ \hline
\end{tabular}
}

\newcommand{\FlaWorksheetSeven}{
\begin{tabular}{| c | p{0.98\textwidth} |}\hline
Step & $\mbox{\color{blue}Algorithm:~}\routinename$
\\ \hline
\rowcolor{yellow!75}
1a & $ \precondition $ 
\\ \whline
4 & 
\begin{minipage}[t]{0.9\textwidth}%
\partitionings~ \\
$\mbox{\color{blue} ~~~where~}$ \partitionsizes
\end{minipage}
\\ \hline
\rowcolor{yellow!75}
2 & $ \invariant $ 
\\ \hline
3 &$\mbox{\color{blue}while~} \guard \mbox{~\color{blue} do}$
\\ \hline 
\rowcolor{yellow!75}
2,3 & ~~~~ $ \invariant \wedge \guard$ 
\\ \hline
5a & ~~~~ \begin{minipage}[t]{0.85\textwidth}%
\ifthenelse{\equal{\blocksize}{1}}{}%
{%
\ifthenelse{ \equal{\blocksize}{blank} }{}%
{{\bf Determine block size $ \blocksize $}\\}%
}
\repartitionings~ \\
$\mbox{\color{blue} ~~~where~}$ \repartitionsizes
\end{minipage}
\\ \hline
\rowcolor{yellow!75}
6 & ~~~~ $\beforeupdate $
\\ \hline
8 & ~~~~  \phantom\update 
\\ \hline
5b & ~~~~ \begin{minipage}[t]{0.85\textwidth}%
\moveboundaries~
\end{minipage}
\\ \hline
\rowcolor{orange!50}    
7 & ~~~~ $\afterupdate $
\\ \hline
\rowcolor{yellow!75}
2 & ~~~~ $ \invariant  $ 
\\ \hline
 &$\mbox{\color{blue} endwhile} $
\\ \hline \whline
\rowcolor{yellow!75}
2 & $ \invariant \wedge \neg( \guard )$ 
\\ \hline
\rowcolor{yellow!75}
1b & $ \postcondition $ 
\\ \hline
\end{tabular}
}

\newcommand{\FlaWorksheetSix}{
\begin{tabular}{| c | p{0.98\textwidth} |}\hline
Step & $\mbox{\color{blue}Algorithm:~}\routinename$
\\ \hline
\rowcolor{yellow!75}
1a & $ \precondition $ 
\\ \whline
4 & 
\begin{minipage}[t]{0.9\textwidth}%
\partitionings~ \\
$\mbox{\color{blue} ~~~where~}$ \partitionsizes
\end{minipage}
\\ \hline
\rowcolor{yellow!75}
2 & $ \invariant $ 
\\ \hline
3 &$\mbox{\color{blue}while~} \guard \mbox{~\color{blue} do}$
\\ \hline 
\rowcolor{yellow!75}
2,3 & ~~~~ $ \invariant \wedge \guard $ 
\\ \hline
5a & ~~~~ \begin{minipage}[t]{0.85\textwidth}%
\ifthenelse{\equal{\blocksize}{1}}{}%
{%
\ifthenelse{ \equal{\blocksize}{blank} }{}%
{{\bf Determine block size $ \blocksize $}\\}%
}
\repartitionings~ \\
$\mbox{\color{blue} ~~~where~}$ \repartitionsizes
\end{minipage}
\\ \hline
\rowcolor{orange!50}   
6 & ~~~~ $\beforeupdate $
\\ \hline
8 & ~~~~  \phantom\update 
\\ \hline
5b & ~~~~ \begin{minipage}[t]{0.85\textwidth}%
\moveboundaries~
\end{minipage}
\\ \hline
\rowcolor{yellow!75}
7 & ~~~~ $\phantom\afterupdate $
\\ \hline
\rowcolor{yellow!75}
2 & ~~~~ $ \invariant  $ 
\\ \hline
 &$\mbox{\color{blue} endwhile} $
\\ \hline \whline
\rowcolor{yellow!75}
2,3 & $ \invariant \wedge \neg( \guard )$ 
\\ \hline
\rowcolor{yellow!75}
1b & $ \postcondition $ 
\\ \hline
\end{tabular}
}

\newcommand{\FlaWorksheetFive}{
\begin{tabular}{| c | p{0.98\textwidth} |}\hline
Step & $\mbox{\color{blue}Algorithm:~}\routinename$
\\ \hline
\rowcolor{yellow!75}
1a & $ \precondition $ 
\\ \whline
4 & 
\begin{minipage}[t]{0.9\textwidth}%
\partitionings~ \\
$\mbox{\color{blue} ~~~where~}$ \partitionsizes
\end{minipage}
\\ \hline
\rowcolor{yellow!75}
2 & $ \invariant $ 
\\ \hline
3 &$\mbox{\color{blue}while~} \guard \mbox{~\color{blue} do}$
\\ \hline 
\rowcolor{yellow!75}
2,3 & ~~~~ $ \invariant \wedge \guard $ 
\\ \hline
\rowcolor{orange!50}   
5a & ~~~~ \begin{minipage}[t]{0.85\textwidth}%
\ifthenelse{\equal{\blocksize}{1}}{}%
{%
\ifthenelse{ \equal{\blocksize}{blank} }{}%
{{\bf Determine block size $ \blocksize $}\\}%
}
\repartitionings~ \\
$\mbox{\color{blue} ~~~where~}$ \repartitionsizes
\end{minipage}
\\ \hline
\rowcolor{yellow!75}
6 & ~~~~ $\phantom\beforeupdate $
\\ \hline
8 & ~~~~  \phantom\update 
\\ \hline
\rowcolor{orange!50}   
5b & ~~~~ \begin{minipage}[t]{0.85\textwidth}%
\moveboundaries~
\end{minipage}
\\ \hline
\rowcolor{yellow!75}
7 & ~~~~ $\phantom\afterupdate $
\\ \hline
\rowcolor{yellow!75}
2 & ~~~~ $ \invariant  $ 
\\ \hline
 &$\mbox{\color{blue} endwhile} $
\\ \hline \whline
\rowcolor{yellow!75}
2,3 & $ \invariant \wedge \neg( \guard )$ 
\\ \hline
\rowcolor{yellow!75}
1b & $ \postcondition $ 
\\ \hline
\end{tabular}
}

\newcommand{\FlaWorksheetFour}{
\begin{tabular}{| c | p{0.98\textwidth} |}\hline
Step & $\mbox{\color{blue}Algorithm:~}\routinename$
\\ \hline
\rowcolor{yellow!75}
1a & $ \precondition $ 
\\ \whline
\rowcolor{orange!50}   
4 & 
\begin{minipage}[t]{0.9\textwidth}%
\partitionings~ \\
$\mbox{\color{blue} ~~~where~}$ \partitionsizes
\end{minipage}
\\ \hline
\rowcolor{yellow!75}
2 & $ \invariant $ 
\\ \hline
3 &$\mbox{\color{blue}while~} \guard \mbox{~\color{blue} do}$
\\ \hline 
\rowcolor{yellow!75}
2,3 & ~~~~ $ \invariant \wedge \guard $ 
\\ \hline
5a & ~~~~ \begin{minipage}[t]{0.85\textwidth}%
\ifthenelse{\equal{\blocksize}{1}}{}%
{%
\ifthenelse{ \equal{\blocksize}{blank} }{}%
{{\bf Determine block size $ \phantom\blocksize $}\\}%
}
$\mbox{\phantom\repartitionings}$~ \\
$\mbox{\color{blue} ~~~where~}$ \phantom\repartitionsizes
\end{minipage}
\\ \hline
\rowcolor{yellow!75}
6 & ~~~~ $\phantom\beforeupdate $
\\ \hline
8 & ~~~~  \phantom\update 
\\ \hline
5b & ~~~~ \begin{minipage}[t]{0.85\textwidth}%
\phantom\moveboundaries~
\end{minipage}
\\ \hline
\rowcolor{yellow!75}
7 & ~~~~ $\phantom\afterupdate $
\\ \hline
\rowcolor{yellow!75}
2 & ~~~~ $ \invariant  $ 
\\ \hline
 &$\mbox{\color{blue} endwhile} $
\\ \hline \whline
\rowcolor{yellow!75}
2,3 & $ \invariant \wedge \neg( \guard )$ 
\\ \hline
\rowcolor{yellow!75}
1b & $ \postcondition $ 
\\ \hline
\end{tabular}
}

\newcommand{\FlaWorksheetThree}{
\begin{tabular}{| c | p{0.98\textwidth} |}\hline
Step & $\mbox{\color{blue}Algorithm:~}\routinename$
\\ \hline
\rowcolor{yellow!75}
1a & $ \precondition $ 
\\ \whline
4 & 
\begin{minipage}[t]{0.9\textwidth}%
$\mbox{\phantom{\partitionings}}$~ \\
$\mbox{\color{blue} ~~~where~}$\phantom{\partitionsizes}  
\end{minipage}
\\ \hline
\rowcolor{yellow!75}
2 & $ \invariant $ 
\\ \hline
\rowcolor{orange!50}  
3 &$\mbox{\color{blue}while~} \guard \mbox{~\color{blue} do}$
\\ \hline 
\rowcolor{orange!50}   
2,3 & ~~~~ $ \invariant \wedge \guard $ 
\\ \hline
5a & ~~~~ \begin{minipage}[t]{0.85\textwidth}%
\ifthenelse{\equal{\blocksize}{1}}{}%
{%
\ifthenelse{ \equal{\blocksize}{blank} }{}%
{{\bf Determine block size $ \phantom\blocksize $}\\}%
}
$\mbox{\phantom\repartitionings}$~ \\
$\mbox{\color{blue} ~~~where~}$ \phantom\repartitionsizes
\end{minipage}
\\ \hline
\rowcolor{yellow!75}
6 & ~~~~ $\phantom\beforeupdate $
\\ \hline
8 & ~~~~  \phantom\update 
\\ \hline
5b & ~~~~ \begin{minipage}[t]{0.85\textwidth}%
\phantom\moveboundaries~
\end{minipage}
\\ \hline
\rowcolor{yellow!75}
7 & ~~~~ $\phantom\afterupdate $
\\ \hline
\rowcolor{yellow!75}
2 & ~~~~ $ \invariant  $ 
\\ \hline
 &$\mbox{\color{blue} endwhile} $
\\ \hline \whline
\rowcolor{orange!50}   
2,3 & $ \invariant \wedge \neg( \guard )$ 
\\ \hline
\rowcolor{yellow!75}
1b & $ \postcondition $ 
\\ \hline
\end{tabular}
}

\newcommand{\FlaWorksheetTwo}{
\begin{tabular}{| c | p{0.98\textwidth} |}\hline
Step & $\mbox{\color{blue}Algorithm:~}\routinename$
\\ \hline
\rowcolor{yellow!75}
1a & $ \precondition $ 
\\ \whline
4 & 
\begin{minipage}[t]{0.9\textwidth}%
$\mbox{\phantom{\partitionings}}$~ \\
$\mbox{\color{blue} ~~~where~} $ \phantom{\partitionsizes} 
\end{minipage}
\\ \hline
\rowcolor{orange!50} 
2 & $ \invariant $ 
\\ \hline
3 &$\mbox{\color{blue}while~} \phantom\guard \mbox{~\color{blue} do}$
\\ \hline 
\rowcolor{orange!50} 
2,3 & ~~~~ $ \invariant \wedge \phantom \guard $ 
\\ \hline
5a & ~~~~ \begin{minipage}[t]{0.85\textwidth}%
\ifthenelse{\equal{\blocksize}{1}}{}%
{%
\ifthenelse{ \equal{\blocksize}{blank} }{}%
{{\bf Determine block size $ \phantom\blocksize $}\\}%
}
$\mbox{\phantom\repartitionings}$~ \\
$\mbox{\color{blue} ~~~where~}$ \phantom\repartitionsizes
\end{minipage}
\\ \hline
\rowcolor{yellow!75}
6 & ~~~~ $\phantom\beforeupdate $
\\ \hline
8 & ~~~~  \phantom\update 
\\ \hline
5b & ~~~~ \begin{minipage}[t]{0.85\textwidth}%
\phantom\moveboundaries~
\end{minipage}
\\ \hline
\rowcolor{yellow!75}
7 & ~~~~ $\phantom\afterupdate $
\\ \hline
\rowcolor{orange!50} 
2 & ~~~~ $ \invariant  $ 
\\ \hline
 &$\mbox{\color{blue} endwhile} $
\\ \hline \whline
\rowcolor{orange!50} 
2 & $ \invariant \wedge \neg( \phantom\guard )$ 
\\ \hline
\rowcolor{yellow!75}
1b & $ \postcondition $ 
\\ \hline
\end{tabular}
}

\newcommand{\FlaWorksheetOne}{
\begin{tabular}{| c | p{0.98\textwidth} |}\hline
Step & $\mbox{\color{blue}Algorithm:~}\routinename$
\\ \hline
\rowcolor{orange!50}
1a & $ \precondition $ 
\\ \whline
4 & 
\begin{minipage}[t]{0.9\textwidth}%
$\mbox{\phantom{\partitionings}}$ ~ \\
$\mbox{\color{blue} ~~~where~}$ \phantom{\partitionsizes} 
\end{minipage}
\\ \hline
\rowcolor{yellow!75}
2 & $ \phantom\invariant $ 
\\ \hline
3 &$\mbox{\color{blue}while~} \phantom\guard \mbox{~\color{blue} do}$
\\ \hline 
\rowcolor{yellow!75}
2,3 & ~~~~ $ \phantom\invariant \wedge \phantom \guard$ 
\\ \hline
5a & ~~~~ \begin{minipage}[t]{0.85\textwidth}%
\ifthenelse{\equal{\blocksize}{1}}{}%
{%
\ifthenelse{ \equal{\blocksize}{blank} }{}%
{{\bf Determine block size $ \phantom\blocksize $}\\}%
}
$\mbox{\phantom\repartitionings}$ ~ \\
$\mbox{\color{blue} ~~~where~}$ \phantom\repartitionsizes
\end{minipage}
\\ \hline
\rowcolor{yellow!75}
6 & ~~~~ $\phantom\beforeupdate $
\\ \hline
8 & ~~~~  \phantom\update 
\\ \hline
5b & ~~~~ \begin{minipage}[t]{0.85\textwidth}%
\phantom\moveboundaries~
\end{minipage}
\\ \hline
\rowcolor{yellow!75}
7 & ~~~~ $\phantom\afterupdate $
\\ \hline
\rowcolor{yellow!75}
2 & ~~~~ $ \phantom\invariant  $ 
\\ \hline
 &$\mbox{\color{blue} endwhile} $
\\ \hline \whline
\rowcolor{yellow!75}
2,3 & $ \phantom\invariant \wedge \neg( \phantom\guard )$ 
\\ \hline
\rowcolor{orange!50}
1b & $ \postcondition $ 
\\ \hline
\end{tabular}
}

\newcommand{\FlaWorksheetZero}{
\begin{tabular}{| c | p{0.98\textwidth} |}\hline
Step & $\mbox{\color{blue}Algorithm:~}\routinename$
\\ \hline
\rowcolor{yellow!75}
1a & $ \phantom\precondition $ 
\\ \whline
4 & 
\begin{minipage}[t]{0.9\textwidth}%
$\mbox{\phantom{\partitionings}}$ ~ \\
$\mbox{\color{blue} ~~~where~}$ \phantom{\partitionsizes} 
\end{minipage}
\\ \hline
\rowcolor{yellow!75}
2 & $ \phantom\invariant $ 
\\ \hline
3 &$\mbox{\color{blue}while~} \phantom\guard \mbox{~\color{blue} do}$
\\ \hline 
\rowcolor{yellow!75}
2,3 & ~~~~ $ \phantom\invariant \wedge \phantom \guard$ 
\\ \hline
5a & ~~~~ \begin{minipage}[t]{0.85\textwidth}%
\ifthenelse{\equal{\blocksize}{1}}{}%
{%
\ifthenelse{ \equal{\blocksize}{blank} }{}%
{{\bf Determine block size $ \phantom\blocksize $}\\}%
}
$\mbox{\phantom\repartitionings}$~ \\
$\mbox{\color{blue} ~~~where~}$ \phantom\repartitionsizes
\end{minipage}
\\ \hline
\rowcolor{yellow!75}
6 & ~~~~ $\phantom\beforeupdate $
\\ \hline
8 & ~~~~  \phantom\update 
\\ \hline
5b & ~~~~ \begin{minipage}[t]{0.85\textwidth}%
\phantom\moveboundaries~
\end{minipage}
\\ \hline
\rowcolor{yellow!75}
7 & ~~~~ $\phantom\afterupdate $
\\ \hline
\rowcolor{yellow!75}
2 & ~~~~ $ \phantom\invariant  $ 
\\ \hline
 &$\mbox{\color{blue} endwhile} $
\\ \hline \whline
\rowcolor{yellow!75}
2,3 & $ \phantom\invariant \wedge \neg( \phantom\guard )$ 
\\ \hline
\rowcolor{yellow!75}
1b & $ \postcondition $ 
\\ \hline
\end{tabular}
}

\newcommand{\TBTinitialize}{}
\newcommand{\FlaAlgorithmTBT}{
\begin{tabular}{|l|} \hline
$\mbox{\color{blue}Algorithm:~}\routinename$
\\ \whline
\partitionings \\
$\mbox{\color{blue} ~~~where~}$ \partitionsizes 
\\ 
\TBTinitialize\\
$\mbox{\color{blue}while~} \guard \mbox{~\color{blue} do}$
\\
\ifthenelse{\equal{\blocksize}{1}}{}%
{%
\ifthenelse{ \equal{\blocksize}{blank} }{}%
{~~~~{\bf Determine block size $ \blocksize $}\\}%
}
~~~~ 
\repartitionings \\
~~~$\mbox{\color{blue} ~~~where~}$ \repartitionsizes
\\ \hline
~~~~  \update 
\\ \hline
~~~~ 
\moveboundaries 
\\
$\mbox{\color{blue} endwhile} $
\\ \hline 
\end{tabular}
}

\renewcommand{\R}{\mathbb{R}}

\newcommand{\CS}[1]{\textrm{CS#1}}
\newcommand{\CA}[1]{\textrm{CA#1}}

\begin{document}


\title{Fast Truncated SVD of 
       Sparse and Dense Matrices on Graphics Processors}





\author{
Andr\'es E. Tom\'as\footnote{Universitat Polit\`ecnica de Val\`encia, Spain.\\
Corresponding author’s email address: antodo@upv.es.} \and
Enrique S. Quintana-Ort\'{\i}\footnote{Universitat Polit\`ecnica de Val\`encia, Spain.} \and
Hartwig Anzt\footnote{Karlsruhe Institute of Technology, Germany; and
               Innovative Computing Laboratory, University of
               Tennessee at Knoxville, USA}
}

\date{June 7, 2023}


\maketitle

\begin{abstract}
We investigate the solution of low-rank matrix approximation problems
using the truncated SVD. For this purpose, we develop and optimize 
GPU implementations for the randomized SVD and
a blocked variant of the Lanczos approach.
Our work takes advantage of the fact
that the two methods are composed of very similar linear
algebra building blocks, which can be assembled using numerical kernels 
from existing high-performance linear algebra libraries. 
Furthermore, the experiments with several sparse
matrices arising in representative real-world applications 
and synthetic dense test matrices reveal a performance advantage 
of the block Lanczos algorithm when targeting the same approximation accuracy.


\end{abstract}

\section{Introduction}

In data science, dimensionality reduction via low-rank matrix approximation is gaining increasing relevance, for example, in order to pre-process large volumes of information prior to the application of machine learning techniques for data synthesis. 
In this line, the \textit{singular value decomposition} (SVD)
is an important low-rank matrix approximation technique, well-known in numerical linear algebra and scientific computing~\cite{GVL3}.

The conventional methods for computing the SVD are quite expensive in terms of floating point arithmetic operations (flops). For this reason, a number
of alternative algorithms have been proposed over the past few years to obtain a low-rank matrix approximation
with a more reduced cost~\cite{Halko11,Martinsson11,10.1137/21M1397866,Yeh22}.
Some of these methods present the additional property that
the accuracy of the approximation can be adjusted by the user.

In this paper, we address the efficient computation 
of low-rank matrix approximations via the computation of a truncated SVD, 
with a special focus on numerical reliability and
high performance, making the following specific contributions:
\begin{itemize}
\item We develop an implementation of the randomized
      SVD method introduced in~\cite{Halko11}, casting its major operations
      in terms of linear algebra building blocks that are especially appropriate
      for data-parallel hardware accelerators such as graphics processing units (GPUs).
\item In addition, we investigate and implement
      an alternative for the truncated SVD based
      on the block Lanczos method~\cite{Golub81}. In doing so, we
      demonstrate that this approach
      can be decomposed into a collection of linear algebra
      building blocks very similar to those in our implementation of
      the randomized SVD algorithm and, therefore, also appropriate for GPUs.
\item We provide a complete numerical evaluation of the two types of methods,
      using a collection of sparse matrices from the Suite Sparse Matrix Collection~\cite{suitesparse}.\\
      At this point, we note that the methods targeted in this work are appropriate
      for both sparse and dense matrices yet the former type of problem is especially interesting. 
      This is possible because, in both types of methods, 
      the problem matrix remains unmodified,
      participating only as an input operand to matrix multiplications.
\item Finally, we complete the experimental analysis of the methods with
      a detailed performance evaluation on 
      an NVIDIA Ampere A100 graphics processor.
\end{itemize}

The rest of the paper is structured as follows.
First, we briefly review the SVD, its use as a tool for obtaining low-rank
approximations, and the two algorithms we consider for this operation.
We then describe in detail the building blocks in these algorithms, and how to customize
them for GPUs. We then evaluate the resulting data-parallel realizations of the
algorithms, from the viewpoints of both numerical accuracy and high performance.
We finally close the paper with some concluding remarks and a discussion of 
future work.
\section{Truncated SVD}
\label{sec:tsvd}

Consider the matrix 
$A \in \mathbb{R}^{m \times n}$
where, without loss of generality, hereafter we assume that $m \ge n$.
(Otherwise, we simply target the transpose of $A$.)
The SVD of the matrix is then given by
\begin{equation}
A = U \Sigma V^T,
\label{eqn:svd}
\end{equation}
where 
$\Sigma = \operatorname{diag}(\sigma_1, \sigma_2, \ldots, \sigma_n)
\in \mathbb{R}^{m \times n}$
is a diagonal matrix containing the singular values
of $A$, while
$U \in \mathbb{R}^{m \times m}$ and
$V \in \mathbb{R}^{n \times n}$ are orthogonal matrices with their
columns respectively corresponding to the left and right singular vectors of the matrix~\cite{GVL3}. 

\newcommand{\ut}{U_{\textsf{T}}}\xspace
\newcommand{\vt}{V_{\textsf{T}}}\xspace
\newcommand{\st}{\Sigma_{\textsf{T}}}\xspace

In many applications, we are interested in obtaining 
a \textit{truncated SVD}, of a certain order $r$,
so that
\begin{equation}
\ut \st \vt^T \approx A,
\label{eqn:tsvd}
\end{equation}
$\st = \operatorname{diag}(\sigma_1, \sigma_2, \ldots, \sigma_r) \in \R^{r \times r}$,
and $\ut,\vt$ contain the first
$r$ columns of $U,V$, respectively.
The practical problem then becomes how to obtain this approximation of $A$
without ``paying the price'' of computing the full decomposition
in~(\ref{eqn:svd}),
which can be considerably higher. This is especially the case when the objective is to obtain a low-rank matrix
approximation, for which $r \ll n$.

We close this short review of the truncated SVD by noting that,
in some cases, the parameter $r$ is not known in advance, but instead
has to be determined based on a user-defined threshold on
the difference
\begin{equation}
\| A - \ut \st \vt^T \|_2 \approx \sigma_{r+1},
\end{equation}
where $\| \cdot \|_2$ denotes the matrix 2-norm.
This leads to the interesting problem of constructing an incremental truncated SVD 
using, for example, an incremental version of the QR factorization~\cite{Gunter:2005:POC}.

In the remainder of this section, we review
two efficient algorithms to compute a truncated SVD:
The randomized SVD and the block Lanczos-based SVD.
These two types can be decomposed 
into a common collection of basic building blocks for matrix
factorizations (Cholesky, QR, SVD), orthogonalization procedures, and
matrix multiplications as described in the next section.

\subsection{Randomized SVD}
\label{subsec:random}

\begin{algorithm}
\caption{RandSVD: Truncated SVD via randomized subspace iteration.}
\label{alg:rsvd}
\centering
\fbox{\begin{minipage}{\columnwidth}
\begin{tabbing}
xxxx\=xx\=xx\=xx\=xx\=\kill
\textbf{Input:} $A \in \mathbb{R}^{m \times n}$; parameters $r \in [1,n]$ and $p,b\geq 1$ \\
\textbf{Output:} $\ut \in \mathbb{R}^{m \times r}, \st = \operatorname{diag}(\sigma_1, \sigma_2, \ldots, \sigma_r),$ \\ 
\> \> $\, \vt \in \mathbb{R}^{n \times r} $ \\ [0.1in]
\>     \' Generate a random matrix $Q_0 \in \mathbb{R}^{n \times r}$ \\
\>     \' \textbf{for} $j = 1,2,\ldots,p$ \\
\>  S1. \' \> $\bar{Y}_j = A Q_{j-1}$ \\
\>  S2. \' \> Factorize $\bar{Y}_j = \bar{Q}_j\bar{R}_j$~~ (Alg.~\ref{alg:cgs-qr}) \\
\>  S3. \' \> $Y_j = A^T \bar{Q}_j$ \\
\>  S4. \' \> Factorize $Y_j = Q_j R_j$~~ (Alg.~\ref{alg:cgs-qr}) \\
\>     \' \textbf{endfor} \\
\>  S5. \' Factorize $R_p = \bar{U} \st \bar{V}^T$~~ (SVD) \\
\>  S6. \' $\ut = \bar{Q}_p \bar{V}$ \\
\>  S7. \' $\vt = Q_p \bar{U}$
\end{tabbing}
\end{minipage}}
\end{algorithm}

\paragraph{Overview.}
The randomized method for the truncated SVD was originally presented 
by~\cite{Martinsson11} and can be derived from Algorithm~\ref{alg:rsvd} by setting $p=1$. 
The idea was subsequently refined in \cite{Halko11} by adding
the subspace iteration to the procedure 
(loop indexed by $p$),
yielding the RandSVD algorithm shown there.

In order to hint why RandSVD delivers a truncated decomposition,  
consider the last iteration of the loop, where $j=p$.
Combining steps~S3 and~S4, we have that
\begin{equation} A^T \bar{Q}_p = Q_p R_p. \end{equation}
Therefore, transposing both sides of the expression and multiplying them
on the left by $\bar{Q}_p$,
\begin{equation} A \approx \bar{Q}_p R_p^T Q_p^T. \end{equation}
Finally, taking into account the SVD in step~S5, 
we obtain that
\begin{equation} 
\begin{array}{rcl}
A &\approx& \bar{Q}_p (\bar{U} \st \bar{V}^T)^T Q_p^T \\ [0.05in]
     &=& (\bar{Q}_p \bar{V}) \st (\bar{U}^T Q_p^T) \\ [0.05in]
     &=& \ut \st \vt^T
\end{array}
\end{equation}
offers the sought-after 
low-rank matrix approximation.

\paragraph{Building blocks.}
From a practical point of view, 
RandSVD comprises
a number of matrix multiplications,
two QR factorizations,
and an SVD. 
We make the following
observations with respect to these operations: 
\begin{itemize}
\item For low-rank matrix approximation problems, $r\ll m,n$. When $A$ is a dense matrix, 
      most of the arithmetic corresponds to the matrix multiplications involving $A$
      (steps S1 and S3).
      The method is suitable for sparse problems because during the iteration $A$ is not modified and, therefore, maintains its sparse structure.
\item Both QR factorizations involve ``tall-and-skinny'' matrices, respectively of dimensions
      $m \times r$ (step~S2) and $n \times r$ (step~S4). Our realization of these factorizations, to be
      presented in the next section, constructs the $Q_j$ and $\bar{Q}_j$ explicitly.
\item After the loop, the SVD (step~S5) 
      operates with a very small ($r \times r$) matrix. 
      The computational cost of 
      this operation is hence negligible.
\item Finally, with the matrices in the sequences $Q_j$ and $\bar{Q}_j$ explicitly built,
      the iteration requires two additional matrix multiplications after the loop (steps S6 and S7).
\end{itemize}

\paragraph{Role of the parameters $p$ and $r$.}
The original RandSVD 
is formulated in our case as a direct method where $p=1$. 
However, this approach may compute very poor approximations of the singular values unless they are well separated. By setting $p>1$, the method performs $p-1$ subspace iterations,
gradually 
improving the accuracy of the computed singular values. In general, a larger value 
for $p$ delivers more accurate approximations. However, as the algorithm exposes,
the computational cost increases linearly with $p$.

The parameter $r$ controls the number of vectors in the subspace iteration and should at least equal the number of singular values that are required.
Typically, $p$ is set to a handful vectors more than the number of singular values to compute.

\subsection{Block Lanczos SVD}
\label{sec:lanczos}

\paragraph{Overview.}
Algorithm~\ref{alg:lsvd}
presents the LancSVD procedure for the truncated SVD based on the 
block Golub-Kahan-Lanczos method~\cite{Golub81}, with the block size parameterized by $b$.
(For simplicity, we assume that $r$ is an integer multiple of $b$.)
The LancSVD algorithm there
is formulated 
with a fixed number of 
iterations,
in order to expose the similarities and differences with RandSVD.


Starting with a random orthonormal matrix $\bar{P_1} \in\mathbb{R}^{m \times b}$, 
at iteration $k$  LancSVD builds 
two matrices, $P_k \in\mathbb{R}^{n \times r}$ and $\bar{P}_k \in\mathbb{R}^{m \times r}$, such that
\begin{equation}
\begin{array}{rcl}
A^T \bar{P}_k &=& P_k B_k, \quad \textrm{and}\\ [0.05in]
A P_k         &=& \bar{P}_k B_k + \bar{Q}_{k+1} R_k E_k,
\end{array}
\end{equation}
where $P_k$ and $\bar{P}_l$ have orthonormal columns, 
(that is, $P_k^T P_k = \bar{P}_k^T \bar{P}_k = I$,
where $I$ denotes the identity matrix of the appropriate order), 
and $\bar{P}_k^T \bar{Q}_{k+1} = I$.
Furthermore, $E_k$ denotes the last $r$ columns 
of an identity matrix of the appropriate order; and
$B_k \in \R^{r \times r}$ is a lower triangular matrix with $b$ non-zero diagonals below the main diagonal
and the following structure:
\begin{equation}
B_k = \left[\begin{array}{ccccc}
                     L_1 \\
                     R_1 & L_2 \\
                         & R_2 & \ddots \\
                         &     & \ddots & L_{k-1} \\
                         &     &        & R_{k-1} & L_k
                     \end{array}\right], 
\end{equation}
where $R_i$ and $L_i$ are respectively upper and lower triangular 
matrices of order $b \times b$.

If the norm of $R_k$ is small, the singular values of $B_k$ approximate the largest $k$ singular values of $A$.
Replacing $B_k$ by its SVD decomposition 
\begin{equation}
B_k = \bar{U} \st \bar{V}^T, 
\end{equation}
we thus obtain 
\begin{equation}
A P_k  = \bar{P}_k B_k + \bar{Q}_{k+1} R_k E_k,
\end{equation}
so that
\begin{equation}
\begin{array}{rcl}
A      &=& \bar{P}_k B_k P_k^T + \bar{Q}_{k+1} R_k E_k P_k^T, \\[0.05in]
       &\approx& \bar{P}_k B_k P_k^T = \bar{P}_k U \st V^T P_k^T.
\end{array}
\end{equation}
The previous equations show also that the left and right singular vectors of $A$ can be obtained from the Lanczos vectors and singular vectors of $B_k$ as follows:
\begin{equation}
\begin{array}{rcl}
U &=& \bar{P}_k \bar{U}, \quad 
V = P_k \bar{V}.
\end{array}
\end{equation}

\begin{algorithm}[t]
\caption{LancSVD: Truncated SVD via block Lanczos method with one-side full orthogonalization and basic restart.}
\label{alg:lsvd}
\centering\fbox{\begin{minipage}{\columnwidth}
\begin{tabbing}
xxxxx\=xx\=xx\=xx\=xx\=xx\=xx\kill
\textbf{Input:} $A \in \mathbb{R}^{m \times n}$; parameters $r \in [1,n]$; $p,b \ge 1$ 
\\
\textbf{Output:} $\ut \in \mathbb{R}^{m \times r}, \st = \operatorname{diag}(\sigma_1, \sigma_2, \ldots, \sigma_r), $ \\
\> $ \vt \in \mathbb{R}^{n \times r} $ \\ [0.1in]
\> S1. \' Generate a random orthonormal matrix\\
\>     \' ~~~~$\bar{Q}_1 \in \mathbb{R}^{m \times b}$ (Alg.~\ref{alg:cholqr2}) \\
\>     \' $k=r/b$ \\
\>     \' \textbf{for} $j = 1,2,\ldots,p$ \\
\>     \' \> \textbf{for} $i = 1,2,\ldots,k$ \\
\> S2. \' \> \> $Q_i = A^T \bar{Q}_{i}$ \\
\>     \' \> \> \textbf{if} $i == 1$ \\
\> S3a. \' \> \> \> Orthogonalize $Q_1$ obtaining $L_1^T$ (Alg.~\ref{alg:cholqr2}) \\
\>     \' \> \> \textbf{else} \\
\> S3b. \' \> \> \> Orthogonalize $Q_i$ against \\
\>     \' \> \> \> \> $P_{i-1} = \left[ Q_1 Q_2 \ldots Q_{i-1} \right]$ \\
\>     \' \> \> \> \> obtaining $H_i$ and $L^T_i$ ~~(Alg.~\ref{alg:cgs-cqr2}) \\
\>     \' \> \> \textbf{endif} \\
 \> S4. \' \> \> $\bar{Q}_{i+1} = A Q_i$ \\
\> S5. \' \> \> Orthogonalize $\bar{Q}_{i+1}$ against \\
\>     \' \> \> \> $\bar{P}_i = \left[ \bar{Q}_1 \bar{Q}_2 \ldots \bar{Q}_i \right]$ \\
\>     \' \> \> \> obtaining $\bar{H}_i$ and $R_i$ ~~(Alg.~\ref{alg:cgs-cqr2}) \\
\>     \' \> \textbf{endfor} \\
\> S6. \' \> Factorize 
             $B_k = 
                     \bar{U} \st \bar{V}^T$ ~~~(SVD)\\ 
\>     \' \> \textbf{if} $j < p$ \\
\>     \' \> \> Split $\bar{U} \rightarrow \left[\bar{U}_1 \bar{U}_2 \ldots \bar{U}_k \right]$ \\ [0.05in]
\> S7. \' \> \> $\bar{Q}_1 = \left[\bar{Q}_1 \bar{Q}_2 \ldots \bar{Q}_k \right] \bar{U}_1 $ \\
\>     \' \> \textbf{endif} \\
\>     \' \textbf{endfor} \\
\> S8. \' $\vt = \left[Q_1 Q_2 \ldots Q_k \right] \bar{V}^T$ \\
\> S9. \' $\ut = \left[\bar{Q}_1 \bar{Q}_2 \ldots \bar{Q}_k \right] \bar{U} $ \\
\end{tabbing}
\end{minipage}}
\end{algorithm}

It is well known that the original Lanczos algorithm implemented in floating point arithmetic fails to compute fully orthogonal matrices. From the multiple solutions proposed in the literature, we choose the full orthogonalization against all previous Lanczos vectors.
This approach is computationally expensive, but it presents the advantage of being composed of large matrix operations, which are very efficient to compute in GPUs.

The main drawback of the full orthogonalization approach is that the computational cost of the Lanczos method rapidly increases with the number of iterations, as each iteration adds new columns to the basis that has to be employed in the orthogonalization.
Also, the amount of memory to store all the previous Lanczos vectors grows linearly.
In order to avoid these issues, a restating technique is frequently used in combination with the Lanczos method.
There are several restarting techniques in the literature,
see for example~\cite{Baglama06}, but for simplicity we choose the original one from~\cite{Golub81}. In this approach, the Lanczos iteration is also run several times, but instead of using random vectors as the initial vectors after each restart, these are set to the approximations of the left singular vectors associated with the $b$ largest singular values. As a result, the new Lanczos iteration maintains the most relevant part of the search directions computed in the previous iteration.

\paragraph{Building blocks.}
We identify the following components in LancSVD, with a significant intersection 
with those present in RandSVD as well as a few differences:
\begin{itemize}
\item  The algorithm comprises 
       matrix multiplications with $A^T$ and $A$ (steps~S2, S4, respectively).
       When $A$ is dense, for low-approximation problems, the arithmetic cost is dominated
       by these matrix multiplications. The algorithm is also appropriate for sparse problems,
       since $A$ is not modified during the computations.
\item The algorithm performs three orthogonalizations 
      (steps S1, S3a/S3b, S4). In the next section
      we will show that the methods for these are akin in our case to that employed for
      the QR factorization present in RandSVD.
\item In the loop there is a small SVD, of size $r \times r$, (step~S5) with a negligible 
      computational cost.
\item Assuming the matrices 
      $Q_1,Q_2,\ldots,Q_k$ and~$\bar{Q}_1,$ $\bar{Q}_2,\ldots,\bar{Q}_k$ 
      are explicitly built,
      there are two additional matrix multiplications after the loop (steps~S8, S9).
\end{itemize}

\paragraph{Role of the parameter $b$.}
Choosing a moderate blocking size $b$ makes the matrix multiplications 
in steps S2, S4 and the orthogonalization in steps S1, S3, S5 more efficient.
Typically, the optimal value for this parameter 
depends on the hardware architecture, 
with the performance initially increasing as it grows, but with a point 
from which the operations do not become any faster.

Furthermore, $b$ should be chosen as large as the number of desired singular values/vectors for maximum effectiveness of the restarting procedure.
In this way, a Lanczos vector is preserved for each wanted singular triplet 
and it is improved at each restart.

\paragraph{Role of the parameter $r$.}

This parameter controls the size of the Krylov subspace generated by LancSVD.
A large value of $r$ improves the convergence, but the cost of the orthogonalization grows at a faster-than-linear pace with it. 
Also, a large amount of memory is required to store all the generated Lanczos vectors.
The convergence rate of the Lanczos procedure mostly depends on the number of matrix applications, which is determined by the ratio $k = r / b$.
When $b=1$, LancSVD becomes the single vector Lanczos iteration with the best convergence rate, but the implementation may be less efficient on a current architecture.

\paragraph{Role of the paramater $p$.}

This parameter allows to continue the Lanczos iteration without incurring in the extra costs of a large $r$.
In a practical implementation of the algorithm, $b$ is set depending on the hardware, $r$ is set taking into account the computation and memory costs, and $p$ is increased till the approximations to the singular triplets satisfy the desired accuracy.

\section{Building Blocks on GPUs}
\label{sec:blocks}

In this section we provide a high level algorithmic description of our realizations 
of the main building blocks identified during the presentation of the RandSVD and LancSVD algorithms. 
In addition, we motivate the selection of the particular realizations chosen for these building blocks, and we
connect (most of) them with high performance implementations of these operations in current
linear algebra libraries for massively data-parallel graphics processors.

\subsection{QR factorization via block Gram-Schmidt}

The loop indexed by $j$ in RandSVD (see Algorithm~\ref{alg:rsvd}) 
comprises the QR factorizations of two tall-and-skinny matrices
per iteration.
These decompositions can be computed, for example, using the
conventional, block column-oriented formulation based on Householder transforms.
Alternatively, due to the dimension of the matrices, it may be more efficient to employ
a communication-avoiding  algorithm for the QR factorization (CAQR) that increases the degree of parallelism~\cite{GVL3,doi:10.1137/080731992,6012824}.

In our case, given that the target architecture is a GPU and,
the triangular factors resulting from the QR factorizations 
in the sequences for $Y_j$ and $\bar{Y}_j$ are not needed (except for the last one),
we decided to compute the factorizations
in RandSVD using a blocked variant of the classical Gram-Schmidt (CGS) method;
see~\cite{doi:10.1137/1.9781611971484}.
This algorithm is simple to implement on GPUs;
it is composed of highly efficient building blocks for this type of data-parallel
hardware accelerators;
and it provides a numerical accuracy similar to that of conventional methods for the QR
factorization for about the same arithmetic cost. 

Algorithm~\ref{alg:cgs-qr} sketches the CGS-QR algorithm for the QR
factorization. 
When called from Algorithm~\ref{alg:rsvd} to factorize the
sequence $Y_j$, the dimension $q=m$. For the sequence $\bar{Y}_j$, $q=n$.
The algorithm basically invokes an 
orthogonalization procedure (steps~S1, S2), using a method to be discussed in the final part of this section.
The remaining operations in the algorithm involve partitioning and data movements
but no relevant arithmetic.



\begin{algorithm}
\caption{CGS-QR: QR factorization via block Gram-Schmidt.}
\label{alg:cgs-qr}
\centering\fbox{\begin{minipage}{\columnwidth}
\begin{tabbing}
xxxx\=xx\=xx\=xx\=xx\=\kill
\textbf{Input:} $ Y \in \mathbb{R}^{q \times r}$; parameter $b \geq 1$ \\
\textbf{Output:} $ Q \in \mathbb{R}^{q \times r}, R \in \mathbb{R}^{r \times r} $ \\ [0.1in]
\>    \' $ k = r / b $ \\ 
\>    \' $Q = Y$, partitioned by column blocks as \\
\>    \> $ Q \rightarrow \left[ Q_1Q_2\ldots Q_k \right]$ \\
\> S1. \' Orthogonalize $Q_1$ obtaining $R_1$ (Alg.~\ref{alg:cholqr2}) \\
\>    \' \textbf{for} $j = 2,3,\ldots,k$ \\
\> S2. \' \> Orthogonalize $Q_j$ against $P_{j-1}= \left[ Q_1 Q_2 \ldots Q_{j-1} \right]$ \\
\> \> \> obtaining $H_j$ and $R_j$ (Alg.~\ref{alg:cgs-cqr2})\\
\>    \' \> Assemble $R_j \leftarrow \left[\begin{array}{c} H_j \\ R_j \end{array}\right]$ \\
\>    \' \textbf{endfor} \\
\>    \' Assemble $R$ by column blocks, $R \leftarrow \left[ R_1 R_2 \ldots R_k \right]$
\end{tabbing}
\end{minipage}}
\end{algorithm}

\subsection{Orthogonalization via CholeskyQR2}

For the orthogonalization procedures required in 
LancSVD and CGS-QR 
(see Algorithms~\ref{alg:lsvd} and~\ref{alg:cgs-qr}, respectively),
we decided to implement a GPU version of 
the CholeskyQR2 algorithm~\cite{CholeskyQR2}. 
In this approach, the procedure is repeated twice to ensure the orthogonality of the resulting 
factor $Q$. In case of a breakdown in the Cholesky decomposition the implementation reverts to a classical Gram-Schmidt with re-orthogonalization.

Algorithm~\ref{alg:cholqr2} shows the CholeskyQR2 orthogonalization procedure.
There, $q=m$ or $q=n$, 
depending on the operation of the algorithm that utilizes it.
The procedure consists of a few basic linear algebra building blocks such
as matrix multiplications
(steps~S1, S4, S7);
triangular system solves (steps~S3, S6); and
two Cholesky factorizations (steps~S2, S5); see~\cite{GVL3}.

\begin{algorithm}
\caption{CholeskyQR2: Orthogonalization.} 
\label{alg:cholqr2}
\centering\fbox{\begin{minipage}{\columnwidth}
\begin{tabbing}
xxxx\=xx\=xx\=xx\=xx\=\kill
\textbf{Input:} $ Q \in \mathbb{R}^{q \times b} $\\
\textbf{Output:} $ Q \in \mathbb{R}^{q \times b}, R \in \mathbb{R}^{b \times b} $ \\ [0.1in]
\> S1. \' $W = Q^T Q$ \\
\> S2. \' Factorize $W = LL^T$  ~~ (Cholesky factorization)\\
\> S3. \' $Q = Q L^{-T}$ \\
\> S4. \' $\bar{W} = Q^T Q$ \\
\> S5. \' Factorize $W = \bar{L}\bar{L}^T$ ~~ (Cholesky factorization) \\
\> S6. \' $Q = Q \bar{L}^{-T}$ \\
\> S7. \' $R = L^T\bar{L}^T$
\end{tabbing}
\end{minipage}}
\end{algorithm}

\subsection{Orthogonalization via CGS and CholeskyQR2}

The complementary orthogonalization procedure,
required by LancSVD and CGS-QR, is implemented using a combination of
block classical Gram-Schmidt and CholeskyQR2. As in the case of the latter, the orthogonalization
procedure performs a second pass, in order to improve the numerical stability of the solution~\cite{10.1007/s00211-005-0615-4}.

Algorithm~\ref{alg:cgs-cqr2} shows the details of this orthogonalization procedure.
When this procedure is called from CGS-QR,
the dimension $q=m$ or $q=n$. 
In both cases, when invoked at iteration $j$ of the corresponding algorithm, 
the dimension $s= (j-1)b$.
The main building blocks in the procedure are the same as those already
identified for CholeskyQR2:
Matrix multiplications
(steps~S1, S2, S3, S6, S7, S8, S11);
triangular system solves (steps~S5, S10); and
two Cholesky factorizations (steps~S4, S9).
There is only one different operation, matrix addition (step S12),
but this contributes a negligible computational cost.

\begin{algorithm}
\caption{CGS-CQR2: Orthogonalization via CGS and CholeskyQR2.}
\label{alg:cgs-cqr2}
\centering\fbox{\begin{minipage}{\columnwidth}
\begin{tabbing}
xxxxx\=xx\=xx\=xx\=xx\=\kill
\textbf{Input:} $ Q \in \mathbb{R}^{q \times b}, P \in \mathbb{R}^{q \times s} $ \\
\textbf{Output:} $ Q \in \mathbb{R}^{q \times b}, H \in \mathbb{R}^{s \times b}, R \in \mathbb{R}^{b \times b} $ \\ [0.1in]
\> S1. \' $H = P^T Q$ \\
\> S2. \' $Q = Q - P H$ \\
\> S3. \' $W = Q^T Q$ \\
\> S4. \' Factorize $W = LL^T$ ~~ (Cholesky factorization)\\
\> S5. \' $Q = Q L^{-T}$ \\
\> S6. \' $\bar{H} = P^T Q$ \\
\> S7. \' $Q = Q - P \bar{H}$ \\
\> S8. \' $W = Q^T Q$ \\
\> S9. \' Factorize $W = \bar{L}\bar{L}^T$ ~~ (Cholesky factorization)\\
\>S10. \' $Q = Q \bar{L}^{-T}$ \\
\>S11. \' $R = L^T \bar{L}^{T}$ \\
\>S12. \' $H = H + \bar{H}$ 
\end{tabbing}
\end{minipage}}
\end{algorithm}

\subsection{Building blocks, high performance libraries, and computational cost}

\begin{table*}
\caption{Algorithms and building blocks considering $A$ is a sparse matrix with $n_z$ nonzero entries. In case the matrix is dense, the operations
with the ``$^*$'' superscript are performed on the GPU using routine GEMM from cuBLAS, and their cost becomes $2mnr$. In the cost expressions, $\CA{3}$, $\CA{4}$, $\CA{5}$ return the cost of algorithms 3 (CGS-QR), 4 (CholeskyQR2) and 5 (CGS-CQR2) as a function of their input parameters. $\CS{1}$, $\CS{2}$, $\CS{3}$,\ldots refer to the costs of the individual steps.}
\label{tab:blocks}
\resizebox{\textwidth}{!}{%
\begin{tabular}{lllllll} \toprule
  Algorithm    & Step & Algorithm/   & Library & Target      & Cost & Matrix \\
               &      & routine      &         &             &      & transfers \\ \midrule
  Alg. 1       & S1   & SpMM$^*$      & cuSPARSE & GPU       & $2n_zr$    & \\
  RandSVD      & S2   & Alg. 3       & --        & Hybrid    & $\CA{3}(b,m,r)$ & \\
               & S3   & SpMM$^*$      & cuSPARSE & GPU       & $2n_zr$    & \\
               & S4   & Alg. 3       & --        & Hybrid    & $\CA{3}(b,n,r)$ & \\
               & S5   & GESVD        & LAPACK    & CPU       & $O(r^3)$   & Transfer $R_p$ GPU$\rightarrow$CPU \\
               & S6   & GEMM         & cuBLAS    & GPU       & $2mr^2$    & Transfer $\bar{V}$~ CPU$\rightarrow$GPU\\
               & S7   & GEMM         & cuBLAS    & GPU       & $2nr^2$    & Transfer $\bar{U}$~ CPU$\rightarrow$GPU \\ \\
               \cline{2-7} ~\\
               & Total cost&              
               \multicolumn{5}{l}{$p\bigg[\CS{1} + \CA{3}(b,m,r) + \CS{3} + \CA{3}(b,n,r)$\bigg]$ + \CS{5} + \CS{6} + \CS{7}$} \\  \\
               \midrule
 Alg. 2        & S1   & Alg. 4       & --       & Hybrid    & $\CA{4}(b,m)$ & \\
 LancSVD       & S2   & SpMM$^*$     & cuSPARSE & GPU       & $2n_zb$    & \\
               & S3a  & Alg. 4       & --       & Hybrid    & $\CA{4}(b,n)$ & \\
               & S3b  & Alg. 5       & --       & Hybrid    & $\CA{5}(b,n,s=(i-1)b)$& \\
               & S4   & SpMM$^*$     & cuSPARSE & GPU       & $2n_zb$    & \\
               & S5   & Alg. 5       & --       & Hybrid    & $\CA{5}(b,m,s=ib)$& \\
               & S6   & GESVD        & LAPACK   & CPU       & $O(r^3)$   & Transfer $B$~ GPU$\rightarrow$CPU \\
               & S7   & GEMM         & cuBLAS   & GPU       & $2bmr$     & Transfer $\bar{U}_1$ CPU$\rightarrow$GPU \\ 
               & S8   & GEMM         & cuBLAS   & GPU       & $2nr^2$    & Transfer $\bar{V}$~ CPU$\rightarrow$GPU \\ 
               & S9   & GEMM         & cuBLAS   & GPU       & $2mr^2$    & Transfer $\bar{U}$~ CPU$\rightarrow$GPU \\ \\
               \cline{2-7} \\
               & Total cost& 
               \multicolumn{5}{l}{$\CS{1} + \sum_{j=1}^{p} \bigg[ \sum_{i=1}^{r/b} \bigg[\CS{2} + \CS{3} + \CS{4} + %
\CA{5}(b,m,ib)\bigg] + \CS{6} + \CS{7}\bigg] + \CS{8} + \CS{9}$} 
               \\ [0.1in]
               & &
               \multicolumn{4}{l}{If $i==1$,~~$\CS{3} = \CA{4}(b,n)$~~else~~$\CS{3} = \CA{5}(b,n,(i-1)b)$}  \\
               & &
               \multicolumn{4}{l}{If $j<p$,~~~~\,$\CS{7} = 2bmr$~~~~~~~~~~else~~$\CS{7} = 0$}  
               \\ \\ \midrule
 Alg. 3        & S1   & Alg. 4       & --        & Hybrid    & $\CA{4}(b,q)$& \\
 CGS-QR        & S2   & Alg. 5       & --        & Hybrid    & $\CA{5}(b,q,s=(j-1)b)$& 
 \\ \\
               \cline{2-7} ~\\
               & Total cost&
               \multicolumn{5}{l}{$\CA{4}(b,q) + \sum_{j=2}^{r/b} \CA{5}(b,q,(j-1)b)$} \\ \\
 \midrule
 Alg. 4        & S1/S4& GEMM         & cuBLAS    & GPU       & $b^2q$     & \\
 CholeskyQR2   & S2/S5& POTRF        & LAPACK    & CPU       & $b^3/3$    & Transfer $W$ GPU$\rightarrow$CPU\\
               & S3/S6& TRSM         & cuBLAS    & GPU       & $b^2q$     & Transfer $L$ CPU$\rightarrow$GPU\\
               & S7   & TRMM         & BLAS      & CPU       & $b^3$      & \\ \\ 
               \cline{2-7} ~\\
               & Total cost& 
               \multicolumn{5}{l}{$\CS{1} + \CS{2} + \cdots + \CS{7}$} \\ \\
               \midrule
 Alg. 5        & S1/S6& GEMM         & cuBLAS    & GPU       & $2bqs$     & \\
 CGS-CQR2      & S2/S7& GEMM         & cuBLAS    & GPU       & $2bqs$     & \\
               & S3/S8& GEMM         & cuBLAS    & GPU       & $2b^2q$    & \\
               & S4/S9& POTRF        & LAPACK    & CPU       & $b^3/3$    & Transfer $W$ GPU$\rightarrow$CPU \\
               &S5/S10& TRSM         & cuBLAS    & GPU       & $b^2q$     & Transfer $L$~ CPU$\rightarrow$GPU \\
               & S11  & TRMM         & BLAS      & CPU       & $b^3$      & \\
               & S12  & ADD          & Custom    & CPU       & $bs$       & \\ \\
               \cline{2-7} ~ \\
               & Total cost&
               \multicolumn{5}{l}{$\CS{1} + \CS{2} + \cdots + \CS{12}$} \\ \\
\end{tabular}}
\end{table*}

\begin{table*}
    \caption{Matrices used in the experiments.}
    \centering
    \begin{tabular}{lrrrlrrr} \toprule
        Matrix & Rows & Columns & $n_z$ &
        Matrix & Rows & Columns & $n_z$\\ \midrule
        12month1 & 12\,471 & 872\,622 & 22\,624\,727 &
ch7-9-b4 & 317\,520 & 105\,840 & 1\,587\,600 \\
ch8-8-b4 & 376\,320 & 117\,600 & 1\,881\,600 &
connectus & 512 & 394\,792 & 1\,127\,525 \\
dbic1 & 43\,200 & 226\,317 & 1\,081\,843 &
degme & 185\,501 & 659\,415 & 8\,127\,528 \\
Delor295K & 295\,734 & 1\,823\,928 & 2\,401\,323 &
Delor338K & 343\,236 & 887\,058 & 4\,211\,599 \\
Delor64K & 64\,719 & 1\,785\,345 & 652\,140 &
ESOC & 327\,062 & 37\,830 & 6\,019\,939 \\
EternityII\_E & 11\,077 & 262\,144 & 1\,503\,732 &
EternityII\_Etilde & 10\,054 & 204\,304 & 1\,170\,516 \\
fome21 & 67\,748 & 216\,350 & 465\,294 &
GL7d15 & 460\,261 & 171\,375 & 6\,080\,381 \\
GL7d16 & 955\,128 & 460\,261 & 14\,488\,881 &
GL7d22 & 349\,443 & 822\,922 & 8\,251\,000 \\
GL7d23 & 105\,054 & 349\,443 & 2\,695\,430 &
Hardesty2 & 929\,901 & 303\,645 & 4\,020\,731 \\
IMDB & 428\,440 & 896\,308 & 3\,782\,463 &
LargeRegFile & 2\,111\,154 & 801\,374 & 4\,944\,201 \\
lp\_nug30 & 52\,260 & 379\,350 & 1\,567\,800 &
lp\_osa\_60 & 10\,280 & 243\,246 & 1\,408\,073 \\
mesh\_deform & 234\,023 & 9\,393 & 853\,829 &
NotreDame\_actors & 392\,400 & 127\,823 & 1\,470\,404 \\
pds-100 & 156\,243 & 514\,577 & 1\,096\,002 &
pds-40 & 66\,844 & 217\,531 & 466\,800 \\
pds-50 & 83\,060 & 275\,814 & 590\,833 &
pds-60 & 99\,431 & 336\,421 & 719\,557 \\
pds-70 & 114\,944 & 390\,005 & 833\,465 &
pds-80 & 129\,181 & 434\,580 & 927\,826 \\
pds-90 & 142\,823 & 475\,448 & 1\,014\,136 &
rail2586 & 2\,586 & 923\,269 & 8\,011\,362 \\
rail4284 & 4\,284 & 1\,096\,894 & 11\,284\,032 &
rel8 & 345\,688 & 12\,347 & 821\,839 \\
rel9 & 9\,888\,048 & 274\,669 & 23\,667\,183 &
relat8 & 345\,688 & 12\,347 & 1\,334\,038 \\
relat9 & 12\,360\,060 & 549\,336 & 38\,955\,420 &
Rucci1 & 1\,977\,885 & 109\,900 & 7\,791\,168 \\
shar\_te2-b2 & 200\,200 & 17\,160 & 600\,600 &
sls & 1\,748\,122 & 62\,729 & 6\,804\,304 \\
spal\_004 & 10\,203 & 321\,696 & 46\,168\,124 &
specular & 477\,976 & 1\,600 & 7\,647\,040 \\
stat96v2 & 29\,089 & 957\,432 & 2\,852\,184 &
stat96v3 & 33\,841 & 1\,113\,780 & 3\,317\,736 \\
stormG2\_1000 & 528\,185 & 1\,377\,306 & 3\,459\,881 &
tp-6 & 142\,752 & 1\,014\,301 & 11\,537\,419 \\ \bottomrule

    \end{tabular}
\end{table*}

Our realizations of RandSVD and LancSVD 
are composed of several basic linear algebra operations such as matrix multiplications
in various forms (sparse, general, triangular) and
triangular system solves, plus two elaborate matrix factorizations (SVD and Cholesky).
To perform the basic operations, we rely on routines from high performance 
linear algebra libraries for GPUs:
cuSPARSE (for sparse matrix multiplications) and 
cuBLAS (for dense and triangular operations).
In the case of the matrix factorizations, due to their complexity and
the small dimension of their input operands, we instead rely on LAPACK to compute them on the CPU.

Table~\ref{tab:blocks}
specifies 1) the routines employed for the operation at each step of the algorithms
and building blocks; 2) the libraries they belong to; 3) the target architecture (GPU,
CPU or hybrid); 4) their theoretical costs (in floating point operations, or flops); and
5) the matrix transfers between CPU and GPU.
The total costs, also reported there, can be derived from the expressions in the table. 
For example, the cost of RandSVD is obtained by adding
the cost of steps S1--S4 multiplied by the number of iterations of the algorithm loop ($p$), 
and next adding also the cost of steps S5--S7. In this case, $\CA{3}$ (acronym for cost of Alg. 3) is a function that returns the cost of 
CholeskyQR2 when invoked with
$r=b$ and either $q=m$ (for step~S2) or $q=m$ (for step~S4).
Similarly,
the cost of CGS-QR is given by
\begin{equation}
\CA{4}(b,q) + \sum_{j=2}^{r/b} \CA{5}(b,q,s),
\end{equation}
where $\CA{4}$, $\CA{5}$ (for cost of Algs. 4, 5) 
are functions that respectively return the cost of CholeskyQR2
and 
the cost of CGS-CQR2 when
invoked at iteration $j$ with $s=(j-1)b$.

Table~\ref{tab:blocks}
exposes that the computational cost
of the truncated SVD algorithm depends
on the problem dimensions ($m,n$ and, for
sparse problems, the number of non-zero elements $n_z$) as well as
on the algorithmic parameters
$p$, $r$ and $b$.
The expression for the total costs
indicate that
the relation between $p$ 
and the cost is linear.
However, the impact of
$r$ on the cost is more subtle, as discussed next:
\begin{itemize}
\item  For RandSVD, increasing $r$ 
will produce linear increments
in the cost for steps S1, S3;
quadratic for S2, S4, S6, S7; and cubic
for S5. 
\item For the LancSVD,
the cost for step S7
grows linearly with $r$;
quadratically for steps S8, S9;
and cubically for step S6.
Increasing $r$ has no impact on an 
individual sparse-general matrix multiplications (SpMM, steps S2 and S4), 
but the number of them
grows linearly with it because we
perform $r/b$ operations of this type. 
Similar reasoning concludes that the
costs of (S3a+S3b) and S4 are quadratic on~$r$.
\end{itemize}
Globally, the impact of $r$ 
on the total
cost is determined by which of these
steps dominate, yet this is
problem-dependent.
Finally, the impact of $b$
on the cost of the two algorithms only affects a few of their steps to a minor extent,
and plays no role in the cost of the
remaining ones.

\section{Experimental results}
\label{sec:result}

The experiments in this section
were carried out, using IEEE double
precision (i.e., 64-bit) arithmetic, 
on a server equipped with
an AMD EPYC 7282 16-core processor (2.8\,GHz) and
504\,GB of DDR4 memory. In addition, the system is 
connected via a PCIe bus to
an NVIDIA A100 graphics
accelerator with 40\,GB of
DDR5 memory.
The platform runs Ubuntu 10.04.6
distribution (Linux Kernel 4.15.0) and 
the following software: \texttt{gcc} v8.4.0, 
CUDA Toolkit 11.2 (including the cuBLAS, cuSPARSE, and cuRAND libraries),
and Intel MKL 2021.2.0.

As suggested in \cite{Martinsson11}, the initial vectors for RandSVD are generated on the GPU using the cuRAND library with a random Poisson distribution with zero mean and deviation of 1. 
The initial vectors for LancSVD are generated in the same manner,
yet with an orthonormalization afterwards (step S1).
All results we report next are averaged over several executions following warm-up runs.


\newcommand{\rr}[1]{{\cal R}_{#1}}\xspace

In the experiments, we
assume that the objective is to 
obtain the 10 largest singular
values/vectors of the problem.
The accuracy of the computed values
will be evaluated using the relative residual
\begin{equation}
\rr{j} = \| Au_j - \sigma_j v_j \|_2 / \sigma_j, 
\end{equation}
where $\sigma_j$, $u_j$ and $v_j$ respectively denote the 
computed
$j$-th singular value, left singular vector and 
right singular vector; and 
$\| \cdot \|_2$ 
stands for the vector 2-norm.
This metric has the advantage of 
combining the reliability of the singular value
and both singular vectors in a single number.


\begin{figure*}
\centering
\includegraphics[width=\textwidth]{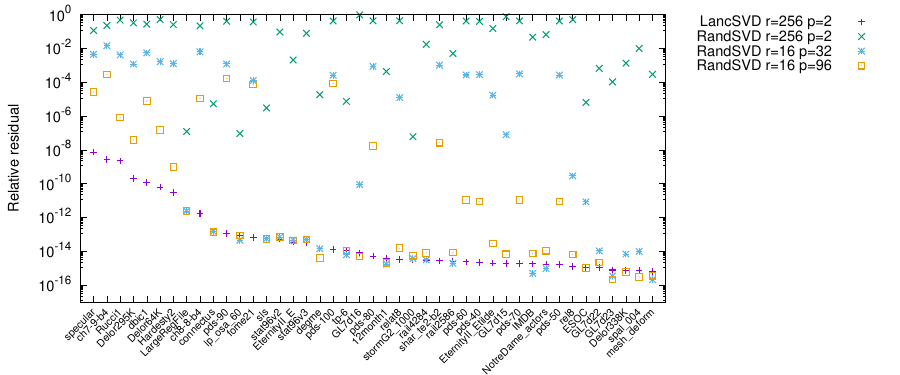}\\
\includegraphics[width=\textwidth]{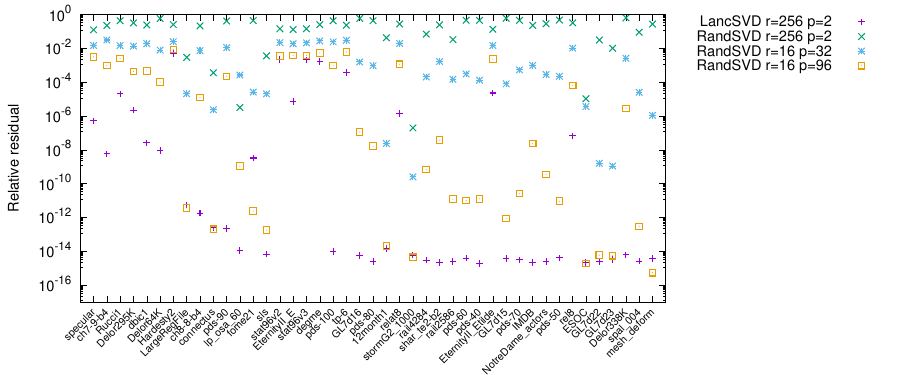}
\caption{Relative residuals $\rr{1}$ (top) and $\rr{10}$ (bottom) 
for the solutions computed with 
RandSVD and LancSVD and different values of $r$ and $p$. In all cases, $b=16$.}
\label{fig:residual}
\end{figure*}

\begin{figure*}
\centering
\includegraphics[width=\textwidth]{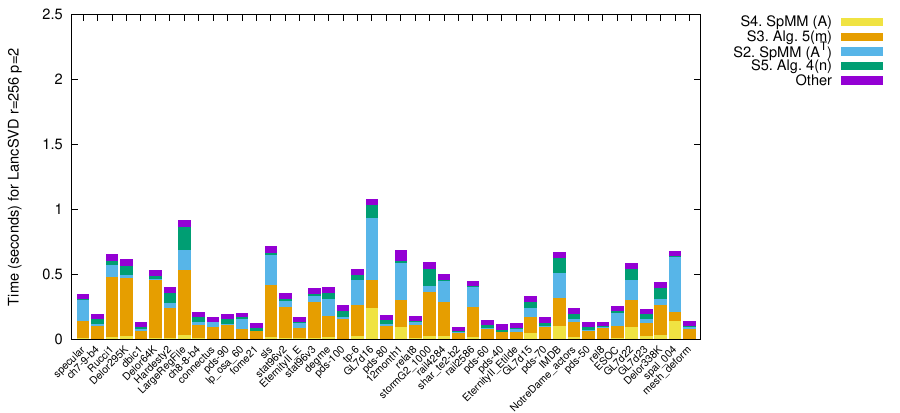}\\
\includegraphics[width=\textwidth]{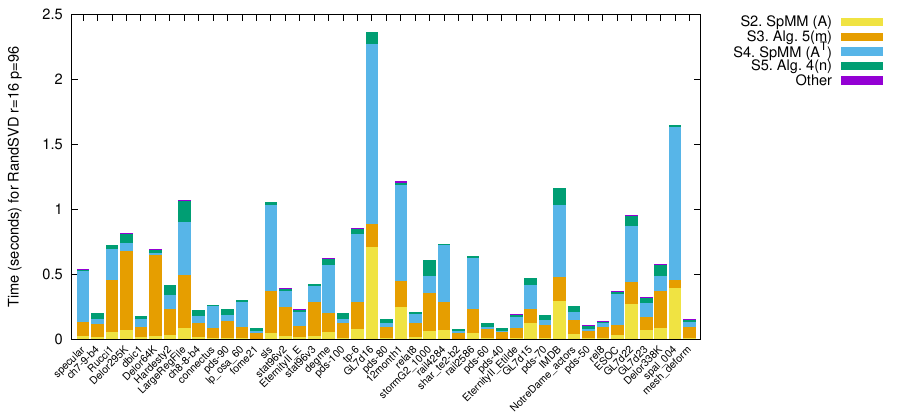}\\
\includegraphics[width=\textwidth]{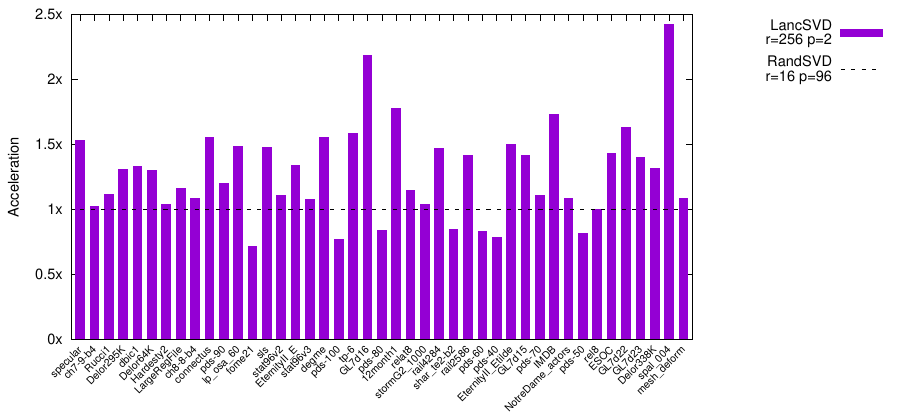}\\
\caption{Execution time of LancSVD and RandSVD (top and middle, respectively) and speed-up of LancSVD with respect to RandSVD (bottom).}
\label{fig:time}
\end{figure*}

\begin{figure*}
\centering
\includegraphics[width=\textwidth]{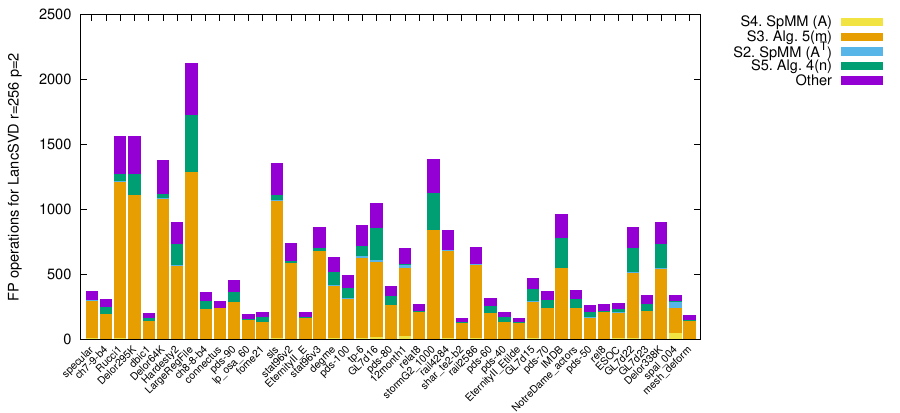}\\
\includegraphics[width=\textwidth]{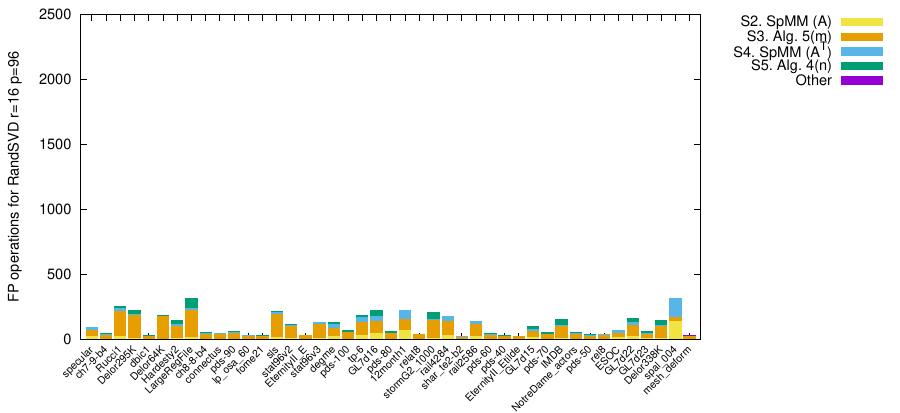}
\caption{Distribution of the flops 
         across the major building blocks in 
         LancSVD and RandSVD (top and bottom, respectively).}
\label{fig:flops}
\end{figure*}


\begin{figure*}
\centering
\includegraphics[width=\textwidth]{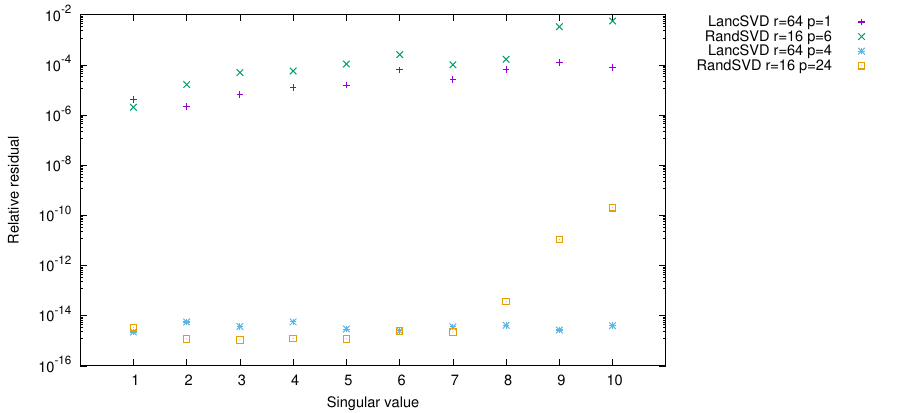}\\
\includegraphics[width=\textwidth]{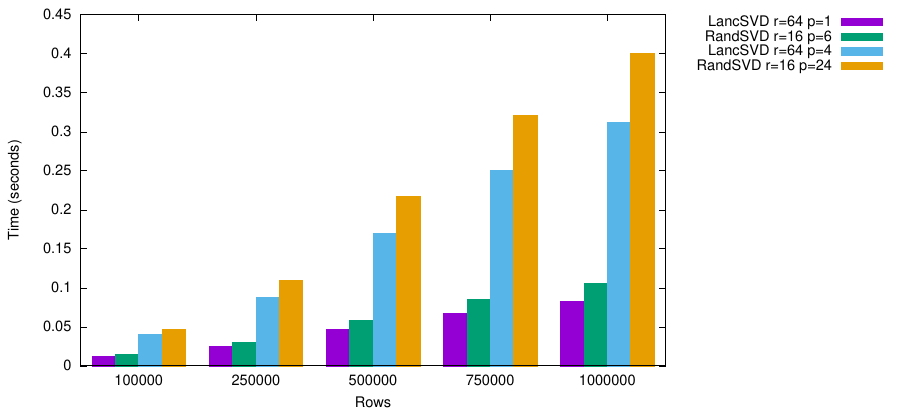}
\caption{Relative residuals $\rr{1}$ to $\rr{10}$ for the solutions computed with the
LancSVD and RandSVD and different values of $r$ and $p$ (top)
and execution time (bottom). In all cases, $b=16$.}
\label{fig:dense}
\end{figure*}


\subsection{Sparse problems}

For the evaluation of the methods with sparse matrices, 
we selected several real cases from
the Suite Sparse Matrix Collection.
In particular, we chose rectangular and large matrices with
more than 200\,000 rows or columns, and the large dimension 
being more than two times larger than the short one.
Matrices are stored in CSR format, using the default cuSPARSE algorithm, as we 
experimentally found that this combination is optimal or sufficiently close to the optimal one for all cases.

\subsubsection{Accuracy}

The discussion at the end of the previous section highlighted the impact of
the algorithmic parameters $p,r,b$ on the
theoretical cost of RandSVD and LancSVD. We next
expose that these parameters exert a key
effect on the convergence rate of the
algorithms and the accuracy of the
computed singular values and vectors.

In order to compute the 10 largest singular values/vectors
with LancSVD, we set $b=16$,  
$r=256$, and $p=2$. 
That is, we perform $r/b=256/16=16$ Lanczos iterations with $p=2$ restarts.
These parameters were determined experimentally by inspecting the computed relative residuals
for different configurations.
This is done to compare LancSVD with RandSVD which, in contrast
with the former, is presented in the literature as a direct algorithm.
In a practical implementation, LancSVD would be formulated as an iterative algorithm where $p$ is increased till the desired accuracy is achieved (within an iteration count limit).

Figure~\ref{fig:residual} 
reveals that the residual
$\rr{1}$ for the solution 
computed with LancSVD is in the range
between $10^{-14}$ and $10^{-8}$; and
between $10^{-14}$ and $10^{-4}$ for 
$\rr{10}$,
except for five cases where the latter lies 
in the range
$[10^{-4},10^{-2}]$.
In the following, we will consider these accuracy levels 
as the baseline for the evaluation of the 
RandSVD counterpart.
For an easy comparison, the results for the sparse problems are
ordered according to decreasing values
of the metric $\rr{1}$ 
for the solution computed with LancSVD.

Figure~\ref{fig:residual}
also shows the accuracy of the solutions
computed with RandSVD for three cases:
\begin{enumerate}
\item $r=256, p=2$, for which the theoretical cost of 
RandSVD is similar to that of LancSVD;
\item $r=16, p=32$, as this case reduces considerably the theoretical
cost of RandSVD compared to that of LancSVD, while maintaining the same number of sparse matrix multiplications;
\item $r=16, p=96$, because these tuple offers a residual $\rr{1}$ 
that is close to that
attained with LancSVD for about half of the cases.
\end{enumerate}
For all these three configurations, we set $b=16$.

The quick conclusion
from this study is that RandSVD requires a large value
of $p$ in order to attain a relative residual that is close 
to that of the solutions computed with LancSVD. 
For that reason, we will only consider RandSVD with $r=16, p=96$ in the following experiments
(and LancSVD with $r=256,p=2$).
We emphasize that, even
with this large value for $p$, the relative
residual of the singular values and vectors 
computed with RandSVD still lags behind that of LancSVD, 
with the difference being larger for the smaller singular values/vectors.
This difference is due to the convergence properties of RandSVD, which are the same as those of the classic subspace iteration method. 
In comparison, LancSVD could exhibit super-linear convergence as is common 
for other Krylov methods~\cite{Simoncini05}.
The lesson to take away is that obtaining an accuracy that is close to the
baseline set by LancSVD may be prohibitively expensive for RandSVD. 


\subsubsection{Performance}
Figure~\ref{fig:time} compares the execution time of the two
algorithms for the truncated SVD. For clarity, the figure also includes
a plot showing the speed-up attained
by LancSVD (with $r=256$, $p=2$) with respect to
RandSVD (with $r=16$, $p=96$).
The results show that LancSVD
is the clear winner, with speed-up factors in the range
$2\times$ to $2.5\times$ 
in two cases; superior to $1.2\times$ for many other
problems; and below $1\times$ for 7 out of 46 matrices only.

The two top plots in 
Figure~\ref{fig:time} also report the distribution of the execution time 
across the
major building blocks in the respective algorithms.
These graphs expose that both algorithms for the truncated SVD spend a significant
portion of the runtime in the SpMM routine with
$A^T$ (step S4 in LancSVD and step S3 in RandSVD)
and the orthogonalization in the $m$-dimension
(step S5 for LancSVD and step S2 for RandSVD).
To put these time costs in perspective, 
we note that the algorithms perform exactly the same number of floating point operations (flops)
in the SpMM with $A$ and $A^T$, showing a clear inefficiency of this
kernel in cuSPARSE when the 
the sparse matrix is to be implicitly transposed as part of the operation and/or
has more columns than rows.
However, explicitly storing a transposed copy of the sparse matrix 
did not yield significant changes in most cases,
and a few ones even exhibited much lower performance due because they present 
a non-zero pattern with some close-to-dense rows.

Figure~\ref{fig:flops}
leverages the analytical cost models 
in Table~\ref{tab:blocks}
to further investigate the actual performance of the building blocks,
showing the distribution of the theoretical flops.
The plots in that figure unveil two interesting issues:
\begin{enumerate}
\item A significant part of flops is necessary for the orthogonalization
      in the $m$-dimension. This is in contrast with the actual execution
      time of this component which, as shown in Figure~\ref{fig:time},
      is less dominant for many cases. Again the clear cause is the slow execution
      of the SpMM kernel when operating with $A^T$.
\item Another striking point is that RandSVD requires fewer
      flops than LancSVD, yet in Figure~\ref{fig:time} 
      the execution time of the former was reported to be considerably larger.
      This is due to the costly SpMM with $A^T$.
      Concretely, LancSVD with $p=2$ performs two products of this
      type. In comparison, 
      RandSVD with $r=16,p=96,b=16$ involves 96 products of the 
      same type. As a result, although RandSVD requires less 
      flops than LancSVD, in practice the low performance of that 
      type of kernel determines the higher execution
      time of the former in most cases.
\end{enumerate}

The results reveal that an improved SpMM algorithm is
the key to augment the performance of both RandSVD and LancSVD. 
This is true in particular for the case where the matrix 
is transposed as the cuSPARSE realization of this building 
block achieves only low performance.
The results also reveal that, in the case of sparse problems, 
LancSVD is superior to RandSVD.
Acknowledging the aforementioned significance of the SpMM kernel 
performance for the RandSVD and LancSVD implementations, 
a question that naturally
arises is whether the LancSVD is \textit{algorithmically}
superior to RandSVD, or whether the superiority is only an artifact
of the implementation of SpMM in cuSPARSE. 
To answer this question, we next investigate the performance of the 
algorithms for dense test matrices.




\subsection{Dense problems}

To compare the reliability and performance of RandSVD and LancSVD for dense matrices,
we designed a synthetic benchmark similar to those leveraged in the literature.
We set the number of columns $n$ to 10\,000 
and the number of rows $m$ to either 100\,000, 250\,000, 750\,000 or 1\,000\,000.
This allows us to explore the effect of a varying number of 
rows on performance.

The dense problems were generated using 
two random orthogonal matrices $X \in\mathbb{R}^{m \times n}, 
Y \in\mathbb{R}^{n \times n}$, 
and setting
\begin{equation}
A = X \Sigma Y^T, 
\end{equation}
where $\Sigma = \textrm{diag}(\sigma_1,\sigma_2,\ldots,\sigma_n) \in \R^{m \times n}$ 
is a diagonal matrix with the desired singular values for the problem.
In the experiments, 
we used
\begin{equation}
\sigma_i = \left\{\begin{array}{ll}
    10^{\frac{15i}{n/2} - 14} & \textrm{if}~~1 \le i \le n/2, \quad \textrm{or} \\
    10^{-14} & \textrm{otherwise}. 
    \end{array}\right.
\end{equation}
This generates a problem where more than half of its singular values are close to zero. 
In addition, the singular values decay asymptotically to the rounding error 
$\epsilon$ in double precision.

As in our previous experiments with sparse matrices, 
we aim to compute 10 singular values, and set $b=16$ as the block size.
We found that, for these dense problems, a smaller size of the subspace, 
with $r=64$, is enough for LancSVD to compute reasonably 
accurate approximations. 
We use two settings where we perform either one ($p=1$) or four ($p=4$)
iterations of LancSVD. 
Using one LancSVD iteration, the approximation accuracy for the singular values ranges between $10^{-6}$ and $10^{-4}$.
The top plot in Figure~\ref{fig:dense} reveals that we need to perform $p=6$ iterations 
of RandSVD \textcolor{black}{with $r=16$} to match this approximation accuracy
If we perform four LancSVD iterations, the approximation accuracy reaches $10^{-14}$. 
We need to perform $p=24$ iterations of RandSVD to match this approximation accuracy.
Thus, in both scenarios, the RandSVD needs \textcolor{black}{roughly} a $6\times$ higher iteration count to match the accuracy
of the LancSVD.
The bottom plot of Figure~\ref{fig:dense} reveals that when aiming for the same approximation quality, 
RandSVD is slower than LancSVD, 
but the speed difference between the two is smaller when the requested accuracy
is low.


\section{Conclusions}
\label{sec:conclusions}

This work illustrates that the randomized algorithm and the block 
Golub-Kahan-Lanczos method for the truncated SVD can be both decomposed
into a number of common basic building blocks. In addition, when the target architecture is a 
massively data-parallel accelerator, such as a GPU,
these building blocks can be directly assembled using 
some orthogonalization procedures based on the Classical Gram-Schmidt \textcolor{black}{and CholeskyQR} algorithms.
The remaining operations in the SVD algorithms comprise a few types of matrix multiplications
and a couple of matrix factorizations with negligible cost.
In summary, current high-performance libraries for GPUs provide the necessary
numerical tools for solving low-rank approximation problems.

A second conclusion of this work is that the computational cost for the
truncated algorithms is dominated, in practice, by that of matrix
multiplications when operating with both dense and sparse problems.
This is particularly painful for the RandSVD algorithm since, in order to obtain
accurate approximations, this method requires a larger number of iterations than
LancSVD and, in consequence, a larger number of matrix multiplications.

Though there may exist scenarios where the RandSVD is superior to LancSVD
\textcolor{black}{(in particular, when the memory capacity of system is limited),}
our experiments indicate that, when targeting the same approximation accuracy,
LancSVD is faster than RandSVD for both dense and sparse problems.

A final remark is that the implementation of 
the sparse matrix multiplication in cuSPARSE, when operating with the transpose
of the sparse matrix, is very slow compared with its non-transposed counterpart. This
has a negative effect on LancSVD and RandSVD, but given the higher number of 
matrix products in the latter, its effect is more profound for that algorithm.

As part of future work, we plan to design
a more sophisticated restart strategy to improve the convergence rate of the
\textcolor{black}{Lanczos} method and, thus, reduce the number of matrix multiplications.
%

\subsection*{Acknowledgements}
This work received funding from
the US Exascale Computing Project (17-SC-20-SC), 
a collaborative effort of the U.S. Department of Energy Office
of Science and the National Nuclear Security Administration;
project PID2020-113656RB-C22 of MCIN/AEI/10.13039/501100011033;
and 
the European High-Performance Computing Joint Undertaking (JU) under grant agreement No
955558 (eFlows4HPC project). The JU receives support from the European Union’s Horizon 2020 research and innovation programme,
and Spain, Germany, France, Italy, Poland, Switzerland, Norway.



\begin{thebibliography}{10}

\bibitem{6012824}
Michael Anderson, Grey Ballard, James Demmel, and Kurt Keutzer.
\newblock Communication-avoiding {QR} decomposition for {GPUs}.
\newblock In {\em 2011 IEEE International Parallel \& Distributed Processing
  Symposium}, pages 48--58, 2011.

\bibitem{Baglama06}
James Baglama and Lothar Reichel.
\newblock Restarted block {Lanczos} bidiagonalization methods.
\newblock {\em Numerical Algorithms}, 43:251--272, 11 2006.

\bibitem{doi:10.1137/1.9781611971484}
\r{A}ke Bj\"{o}rck.
\newblock {\em Numerical Methods for Least Squares Problems}.
\newblock Society for Industrial and Applied Mathematics, 1996.

\bibitem{suitesparse}
Timothy~A. Davis and Yifan Hu.
\newblock {The University of Florida} sparse matrix collection.
\newblock {\em ACM Transactions on Mathematical Software}, 38, 11 2011.

\bibitem{doi:10.1137/080731992}
James Demmel, Laura Grigori, Mark Hoemmen, and Julien Langou.
\newblock Communication-optimal parallel and sequential {QR} and {LU}
  factorizations.
\newblock {\em SIAM Journal on Scientific Computing}, 34(1):A206--A239, 2012.

\bibitem{CholeskyQR2}
Takeshi Fukaya, Yuji Nakatsukasa, Yuka Yanagisawa, and Yusaku Yamamoto.
\newblock {CholeskyQR2}: A simple and communication-avoiding algorithm for
  computing a tall-skinny {QR} factorization on a large-scale parallel system.
\newblock In {\em 2014 5th Workshop on Latest Advances in Scalable Algorithms
  for Large-Scale Systems}, pages 31--38, 2014.

\bibitem{10.1007/s00211-005-0615-4}
Luc Giraud, Julien Langou, Miroslav Rozlo\v{z}n\'{\i}k, and Jasper van~den
  Eshof.
\newblock Rounding error analysis of the {Classical Gram-Schmidt}
  orthogonalization process.
\newblock {\em Numer. Math.}, 101(1):87–100, jul 2005.

\bibitem{GVL3}
Gene~H. Golub and Charles F.~Van Loan.
\newblock {\em Matrix Computations}.
\newblock The Johns Hopkins University Press, Baltimore, 3rd edition, 1996.

\bibitem{Golub81}
Gene~H. Golub, Franklin~T. Luk, and Michael~L. Overton.
\newblock A block {Lanczos} method for computing the singular values and
  corresponding singular vectors of a matrix.
\newblock {\em ACM Trans. Math. Softw.}, 7(2):149–169, jun 1981.

\bibitem{Gunter:2005:POC}
Brian~C. Gunter and Robert~A. van~de Geijn.
\newblock Parallel out-of-core computation and updating the {QR} factorization.
\newblock {\em ACM Trans. Math. Soft.}, 31(1):60--78, March 2005.

\bibitem{Halko11}
N.~Halko, P.~G. Martinsson, and J.~A. Tropp.
\newblock Finding structure with randomness: Probabilistic algorithms for
  constructing approximate matrix decompositions.
\newblock {\em SIAM Review}, 53(2):217--288, 2011.

\bibitem{10.1137/21M1397866}
Eric Hallman.
\newblock A block bidiagonalization method for fixed-accuracy low-rank matrix
  approximation.
\newblock {\em SIAM J. Matrix Anal. Appl.}, 43(2):661–680, jan 2022.

\bibitem{Martinsson11}
Per-Gunnar Martinsson, Vladimir Rokhlin, and Mark Tygert.
\newblock A randomized algorithm for the decomposition of matrices.
\newblock {\em Applied and Computational Harmonic Analysis}, 30(1):47--68,
  2011.

\bibitem{Simoncini05}
Valeria Simoncini and Daniel~B. Szyld.
\newblock On the occurrence of superlinear convergence of exact and inexact
  {Krylov} subspace methods.
\newblock {\em SIAM Review}, 47(2):247--272, 2005.

\bibitem{Yeh22}
Michael Yen.
\newblock {\em An Efficient, Tolerance-Based Algorithm for the Truncated
  {SVD}}.
\newblock PhD thesis, UC Berkeley, USA, 2022.

\end{thebibliography}

\end{document}